# Unique Itinerant Ferromagnetism in $4d$-electron System $Ca_2RuO_4$

Fumihiko NAKAMURA[1*], Hiroki OGURA[1], Tatsuhiro SAKAMI[1], and Takashi SUZUKI[2]

[1]*Kurume Institute of Technology, 2228-66 Kamitsu, Kurume 830-0052, Japan*
[2]*Department of Quantum Matter, AdSE, Hiroshima University, Higashi-Hiroshima 739-8530, Japan*

*E-mail: fumihiko@kurume-it.ac.jp*

We have studied the magnetic properties of pressure-induced ferromagnet $Ca_2RuO_4$ to reveal the uniqueness of the $4d$-electron ferromagnetism in the quasi-two-dimensional conductor. The magnetic parameters have been estimated from the paramagnetic susceptibility and the magnetisation process under pressure up to ~2 GPa. The parameters can well be interpreted on the basis of the self-consistent renormalization theory of spin fluctuation for "three-dimensional" itinerant ferromagnet. Nevertheless, the metallic $Ca_2RuO_4$ shows quite strong anisotropy not only in the conductivity but also in the magnetisation process. Such the strong anisotropy is rare for an itinerant ferromagnet and is a unique characteristic of the $4d$ electron system $Ca_2RuO_4$.

## 1. Introduction

Recently, the "quantum phase transition" has been proposed as a key concept for the unconventional superconductivity in the vicinity of magnetic ordered states [1, 2]. In particular, much attention has recently been gained on the unconventional super-conductivity in the vicinity of a ferromagnetic (FM) state while it is generally known ferromagnetism (FM) plays as a competitive factor for superconductivity. Moreover, we can fully expect unconventional superconductivity due to FM fluctuations from analogy with the triplet pairing in the superfluid 3He [3].

Another interest is a role of two-dimensional (2D) electronic state in a FM magnetic ordering. Although low dimensionality is generally known as a destructive factor for long-range order, exotic quantum phenomena have often been found in quasi two-dimensional (Q2D) systems such as high-$T_c$ cuprates. As seen in previous report, SC-$T_c$ in cuprate superconductors can be risen by releasing frustration due to the 2D electronic states [4]. Now, our concern has been focused on how about ferromagnetism. There is, however, a great lack of experimental studies of the itinerant ferromagnetism in a Q2D metal. To our knowledge, the pressurised $Ca_2RuO_4$ (CRO) is one of the most suitable systems to investigate the relation between itinerant FM and unconventional superconductivity in a quasi-two-dimensional (Q2D) metal.

A single-layered ruthenate ($Ru^{4+}$-$4d^4$) CRO is a Mott insulator and a compound isostructural with the exotic superconductor $Sr_2RuO_4$ [5]. The pressurised CRO is a rare system that the FM order actually occurs in a Q2D metal with a strongly anisotropic conduction [6]. Application of pressure (*P*) to CRO induces versatile quantum phenomena, ranging from the insulator to a superconductor (above ~9 GPa with maximum SC-$T_c$ ~0.4 K) via a Q2D metal with a FM ground state (in the *P* range from 0.5 to 8 GPa with maximum FM-$T_C$ ~25 K) [6, 7]. The FM in the pressurised CRO is, thus, attractive system; however, the quantitative characteristic of the itinerant FM remains unknown so far.

In this report, we mention the magnetisation process and the paramagnetic susceptibility of a Q2D metal of CRO under *P* up to ~2 GPa in order to grasp the FM parameters such as the anisotropy and the itinerancy. In particular, the itinerant FM parameters are quantitatively evaluated on the basis of the self-consistent renormalization (SCR) theory of spin fluctuation for itinerant ferromagnet [8].

## 2. Experiments

In order to evaluate the FM parameters in pressurised CRO, temperature (*T*) and magnetic field ($\mu_0H$) dependences of the magnetisation (*M*) under *P* up to ~2 GPa were measured by using our developed cell [9] equipped with a commercial SQUID magnetometer (Quantum Design, model MPMS). We used the Daphne oil 7243 (Idemitsu Kosan Co., Ltd.) as a pressure transmitting medium. Pressure at low temperatures was estimated by the suppression of SC-$T_c$ of tin loaded within the *P* cell. Our measurement was performed by using several pieces of single-crystalline CRO (total masses ~20 mg). Our crystals grown by a floating-zone method are of single phase of $K_2NiF_4$ structure with the *c* axis of 11.915 Å. Thus, our single crystals are high-purity with essentially stoichiometric oxygen content.

Figure 1(a) shows the comparison among the magnetisation curves at 2 K and 1.8 GPa as a function of the fields along the *a*, *b* and *c* axes up to $\mu_0H$ =5.5 T. It can clearly be seen that the *P*-induced FM in CRO is strongly anisotropic. We note here that the *a* axis is the easy direction of the magnetisation and the *c* axis is the hard direction. Moreover, we deduce that application of $\mu_0 H_{\|c}$ above the anisotropy field ($\mu_0 H_A$) forces the direction of the spin orientation from the *a* to the *c* axis. Indeed, the linearly extrapolated magnetisations of $M_{\|a}$, $M_{\|b}$ and $M_{\|c}$ cross at the field of ~9.5 T, at which the transverse magnetoresistance in $\mu_0 H_{\|c}$ peaks as shown in Fig. 1(b) (The magnetoresistance in $\mu_0 H_{\|c}$ is partly taken from Ref. 10). From the extrapolation, we obtain the anisotropy field to be $\mu_0 H_A$ ~9.5 T.

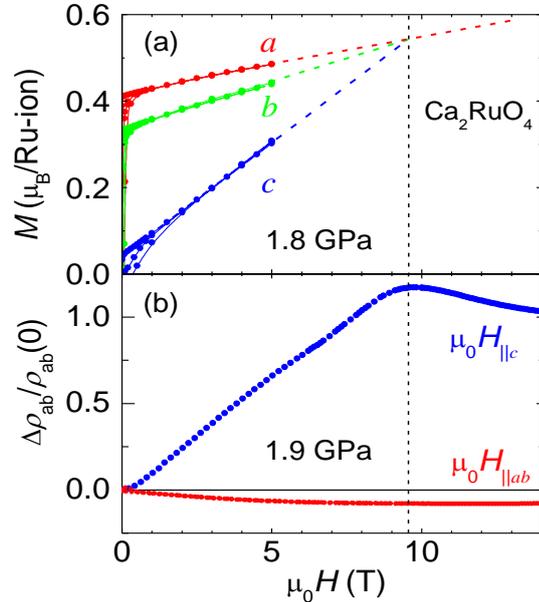

Fig. 1. Magnetic field variations of (a) magnetisation curves in the fields along the *a*, *b* and *c* and (b) transverse magnetoresistance in the fields along the *ab* and the *c* axes. Linearly extrapolated $M_{||a}$, $M_{||b}$ and $M_{||c}$ cross at the field of ~9.5 T. They were measured at 2 K and ~2 GPa.

In order to evaluate the itinerant FM parameters, it is important to measure the magnetisation in the field along the easy direction (namely, the *a*-axis). Figure. 2 (a) and (b) shows the remnant FM moment ($M_{rem}$) at 2 K and the effective paramagnetic (PM) moment ($p_{eff}$) as a function of *P*, respectively. The $M_{rem}$ was obtained from the magnetisation curve in the field of $\mu_0 H_{\|a}$ at 2 K. The FM moment was induced by pressurising over 0.5 GPa. The value of $M_{rem}$ increases rapidly from ~0.1 $\mu_B$/Ru-ion at 0.6 GPa but saturates toward ~0.44 $\mu_B$/Ru-ion pressuring over ~1.7 GPa. The inset of (b) representatively shows inverse PM susceptibility ($\mu_0 H_{\|a}/M$) at 2.0 GPa as a function of *T*. We observed a linear relation obeying the Curie-Weiss law $M/\mu_0 H_{\|a}= C/(T-\theta_p)$. We estimated the Curie temperature $\theta_p$ ~ +10 K and the effective PM moment $p_{eff}$ ~1.85 $\mu_B$. Moreover, these parameters are almost constant in the *P* range above 1.5 GPa.

Next, we mention the itinerancy of the *P*-induced FM in CRO, focusing on the following four viewpoints: First, the FM magnetisation linearly increases without saturation even in strong fields. Second, $M_{rem}$~0.44 $\mu_B$/Ru-ion is much smaller than the saturated moment of 2 $\mu_B$/Ru-ion as a localised system of *S*=1. Third, although a Curie-Weiss type susceptibility has been observed above FM-$T_C$ ~ +10 K, the value of $p_{eff}$ ~1.85 $\mu_B$/Ru-ion is quite smaller than 2.73 $\mu_B$ estimated as a localised system of *S*=1. Last, a ratio of $p_{eff}/M_{rem}$ ~4.2 is much larger than 1 of localised spins systems. It can, therefore, be seen that the *P*-induced FM nature in CRO is quantitatively interpreted in terms of an itinerant FM.

Moreover, we have observed that the *c*-axis susceptibility also obeys the Curie-Weiss law with the parameters of $\theta_p$ ~ −15 K and $p_{eff}$ ~2.1 $\mu_B$ at 2 GPa. Thus, the PM susceptibility is quantitatively anisotropic between the *a* and the *c* axes.

In order to quantitatively evaluate the itinerant parameters, the magnetisation process has been interpreted on the basis of the SCR

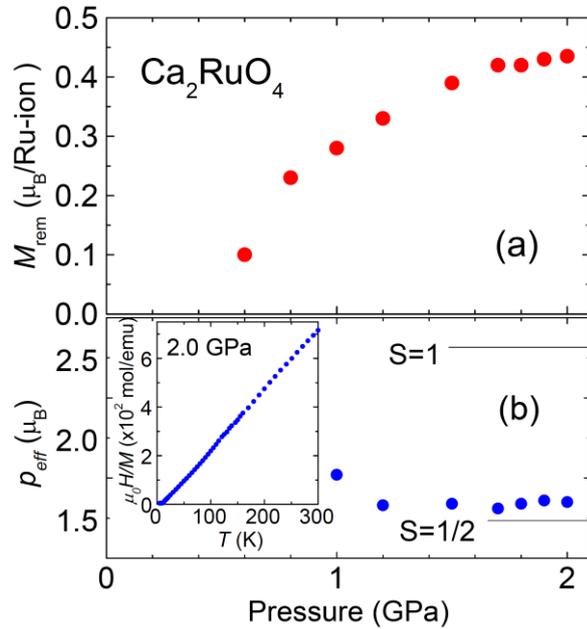

Fig. 2. Pressure variation of the FM parameters estimated from the magnetisation in the pressurised CRO. (a) Remnant magnetisation at 2 K. (b) Effective PM moment estimated from a Curie constant. Inverse susceptibility at 2.0 GPa is representatively plotted as a function of temperature in inset (b).

theory for three-dimensional (3D) spin fluctuations. We, initially, refer to the Arrott's plot criterion [8]. That is, the square of $M$ at 2.0 GPa for several-fixed $T$ is plotted as a function of $\mu_0H/M$ as shown in Fig. 3(a). The magnetisation process at the lowest temperature of 2 K (namely ground state) shows good linearity in the Arrott's plot. From its slope, we have estimated two energy scales of $T_0 \sim 80$ K and $T_A \sim 550$ K, where $T_0$ and $T_A$ characterize the spectral distribution of the spin-fluctuation spectrum in the frequency and wave-vector spaces, respectively.

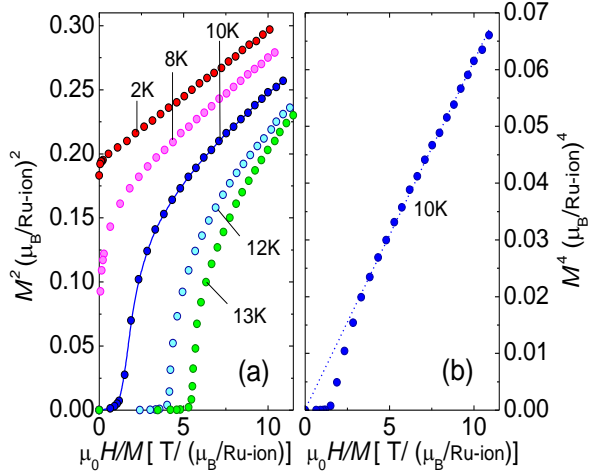

**Fig. 3.** Representative magnetisation curves at 2.0 GPa analysed by the Arrott-plot criterion. (a) The square of the magnetisation (from Ref. 12) is plotted as a function of $\mu_0H/M$. (b) The fourth power of the magnetisation $T_C \sim 10$ K is replotted as a function of $\mu_0H/M$.

A number of experimental and theoretical works indicate that a typical magnetisation process is characterised by a linear relation in the Arrott's plot, especially the proportional relation at FM-$T_C$. The pressurised CRO, indeed, shows the good linear relation at 2 K. However, the magnetisation process in the vicinity of FM-$T_C$ shows a proportional relation not in the $M^2$–$\mu_0H/M$ plot but in the $M^4$–$\mu_0H/M$ plot, as shown in Fig. 3(b). Similar behaviour has been reported in the system with the specific-range parameters of $p_{eff}/M_{rem}$ and $T_C/T_0$ [8]. This is due to that thermal spin fluctuations cannot be neglect at the critical temperature in such the system. We expect that itinerancy of the $P$-induced FM in CRO is close to that in MnSi [8].

Let us compare the itinerancy of the FM in CRO against other ferromagnets reported so far. The obtained parameters are additionally plotted in the generalized Rhodes-Wohlfarth plot for itinerant ferromagnets [8, 11, 12] as shown in Fig. 4, where $p_{eff}/M_{rem}$ is plotted as a function of $T_C/T_0$. The itinerant parameters of the FM in CRO in the $P$ range from 1.0 to 2.0 GPa are well located on the theoretical curve (see also in the inset). It can be seen that the pressurised CRO is an itinerant ferromagnet. Moreover, itinerancy of the CRO is very close to that of the above-mentioned system "MnSi".

Thus, itinerant nature of the $P$-induced FM in CRO can well be interpreted on the basis of the SCR theory for "three-dimensional" (3D) spin fluctuations although the conduction shows essentially 2D metal. Now, we consider how about role of "2D" in itinerant ferromagnet of the pressurised CRO. As shown in our previous report, the in-plane resistance shows metallic behaviour whereas the $c$-axis resistance has a negative slope indicating non-metallic conduction [4]. That is, the system should be interpreted as not a 3D metal with strong anisotropy but an essentially 2D metal. On the other hand, we indicate the FM parameters in pressurised CRO can be explained by the SCR theory for a "3D" spin system rather than "2D" one. It may be difficult to realize a 2D spin system comparing with theoretical predictions even in an ideal 2D conductor such as pressurised CRO. This might be because the ferromagnetic interlayer interaction is hard to suppress even if the conduction among RuO$_2$ planes can electrically be disturbed.

Move on last topic of the FM anisotropy in the pressurised CRO, namely a role of spin-orbit coupling in the magnetism. As shown in previous reports, spin-orbit coupling is a key to understand the attractive phenomena, such as the orbital ordering [13, 14], the giant magnetoresistance [10] and the insulator-metal transition accompanied by a bulk structural transition [15-17]. We note here that FM anisotropy is known as an indicator of the strongness of spin-orbit coupling. Nevertheless, there have been few discussions of a role of spin-orbit coupling in a FM ordering. This is because many of studies for itinerant FM have been concentrated to a 3$d$-electron metallic system where the orbital angular momentum is quenched, and then there have been few reports on a role of spin-orbital coupling in a $d$-electron FM system. It is actually known that the FM anisotropy is quite small in many of 3$d$-electron systems, except for the cobalt metal, where the orbital quenching is most likely incomplete [18, 19].

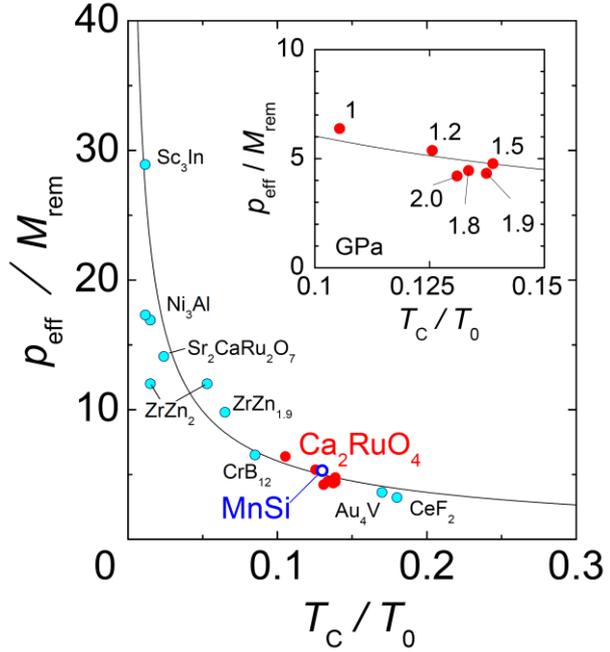

**Fig. 4.** The itinerant FM parameters of pressurised CRO are plotted in the generalized Rhodes-Wohlfarth plot after Ref. 8, 11 and 12.

On the other hand, strong anisotropy has been reported in some uranium FMs (100-1000 time larger than in 3$d$ FMs) [20]. For example, the itinerancy of uranium compounds URhSi and URhGe is comparable to that of the pressurised CRO although their FM anisotropy is quite stronger than that of the pressurised CRO. Moreover, the magnetic and conductive properties in uranium compounds are mainly governed by 5$f$ electrons having intermediate natures between localized 4$f$ and itinerant 3$d$ electrons. We note that strong spin-orbit coupling is a key character of 5$f$ electrons because the orbital angular momentum of 5$f$ electrons is not quenched.

As mentioned above, the $P$-induced FM in 4$d$-electron system CRO is characterised by strong anisotropy as a $d$-electron FM. The anisotropic energy of $E_A \sim 2.1 \times 10^5$ J/m$^3$ was obtained by using the experimental values of $\mu_0 H_A \sim 9.5$ T and $M_{rem} \sim 0.44$ $\mu_B$/Ru-ion [19]. This result indicates that quenching the orbital angular momentum in 4$d$ electron system of CRO is not completely.

This deduction is also indicated by some phenomena such as giant magneto-resistance [10], the orbital-selective Mott transition [21, 22], orbital ordering [13, 14], and the Mott insulator-metal switching induced by a quite small electric-field [23]. All of them strongly suggest that the FM properties in CRO are quite sensitive to strong spin-orbit coupling. Namely, many of interesting phenomena in CRO can be interpreted in terms that the orbital angular momentum is not entirely quenching in 4$d$ electron.

## 3. Conclusion

We have demonstrated the uniqueness of the *P*-induced FM in CRO. First, the itinerancy of the FM is comparable to that in the typical 3D itinerant-ferromagnet MnSi. Second, it is well interpreted in terms of the SCR theory for a "3D" spin system although the pressurised CRO is an essentially 2D metal in electric conduction. Last, we have observed quite strongly anisotropy in the itinerant FM as a *d*-electron system. Such a strong anisotropy in FM is difficult to understand regardless of considering the strong spin-orbit coupling. We, thus, expect to build up a theory of itinerant FM including the strong spin-orbit coupling.

## Acknowledgment

We acknowledge T. Takemoto, R. Nakai and Y. Kimura for their experimental helps. A part of this work was supported by Grant-in-Aid for Scientific Research (Grant Nos. 26247060, 26287083, 17H06136, 22K03485 and 22H01166) by JSPS.